\documentclass[preprint]{aastex}
\pdfoutput=1
\usepackage{times}
\usepackage{wasysym} 
\usepackage{caption} 
\usepackage{subcaption} 
\usepackage{hyperref} 
\usepackage{natbib} 
\usepackage{multirow}

\let\oldtabular\tabular
\renewcommand{\tabular}{\scriptsize\oldtabular}

\usepackage[normalem]{ulem}

\begin{document}
\title{Transiting Planets with LSST - I. Potential for LSST Exoplanet Detection}
\author{Michael B. Lund\altaffilmark{1},
Joshua Pepper\altaffilmark{2,1},
Keivan G. Stassun\altaffilmark{1,3}}
\altaffiltext{1}{Department of Physics and Astronomy, Vanderbilt University, Nashville, TN 37235, USA; \\ \url{michael.b.lund@vanderbilt.edu}}
\altaffiltext{2}{Department of Physics, Lehigh University, Bethlehem, PA 18015, USA}
\altaffiltext{3}{Department of Physics, Fisk University, Nashville, TN 37208, USA}

\captionsetup[table]{labelsep=space}
\captionsetup[figure]{labelsep=space}
\begin{abstract}
The Large Synoptic Survey Telescope (LSST) is designed to meet several scientific objectives over a ten-year synoptic sky survey. Beyond its primary goals, the large amount of LSST data can be exploited for additional scientific purposes. We show that LSST data are sufficient to detect the transits of exoplanets, including planets orbiting stars that are members of stellar populations that have so far been largely unexplored. Using simulated LSST lightcurves, we find that existing transit detection algorithms can identify the signatures of Hot Jupiters around solar-type stars, Hot Neptunes around K-dwarfs, and (in favorable cases) Super-Earths in habitable-zone orbits of M-dwarfs. We also find that LSST may identify Hot Jupiters orbiting stars in the Large Magellanic Cloud---a remarkable possibility that would advance exoplanet science into the extragalactic regime.
\end{abstract}
\emph{Subject headings:} planetary systems --- planets and satellites: detection --- surveys

\section{Introduction}
The ongoing search for exoplanets has to date yielded a sample of some 1,500 planets, primarily from the contributions of the \emph{Kepler} mission \citep{Batalha2012}, ground-based transit surveys, such as HATnet \citep{Bakos2011}, TrES \citep{O'Donovan2007}, SuperWASP \citep{Doyle2012}, XO \citep{Poleski2010}, KELT \citep{Pepper2007}, and RV surveys, such as the Anglo-Australian Planet Search \citep{Tinney2001}, HARPS \citep{Zechmeister2013}, California and Carnegie Planet Search \citep{Patel2007}, and the Lick Planet Search \citep{Fischer2014}. However the current sample predominantly comprises host stars that are F, G, and K dwarfs in the solar neighborhood. Due to a variety of observational and selection biases, other populations have not been examined as thoroughly, including late-type stars, evolved stars, stars outside the solar neighborhood, young stars, stellar remnants, and others. As a consequence, while the current sample of planets has allowed us to characterize exoplanet frequency as a function of mass, radius, and period \citep{Howard2012,Petigura2013,Youdin2011}, the current exoplanet sample does not include stellar hosts that are fully representative of all possiblities.

In this paper, we introduce an analytic- and simulation-based effort to explore the ability of the upcoming Large Synoptic Survey Telescope \citep{LSSTScience2009} to discover transiting exoplanets, specifically among distant stellar populations and late-type stars. Since LSST was not specifically designed for this purpose, there are at least two important difficulties involved. The two most prominent are: (1) The stars observed by LSST are too faint for candidates to be confirmed as exoplanets through standard observational methods, making it difficult to separate real exoplanets from the many types of false positives, and (2) the cadence of LSST observations is much lower than current transit surveys, making robust detection of transit-like events against background noise more difficult. On the other hand, LSST will observe exponentially more stars than prior surveys, and it is prudent to examine just what kinds of exoplanets can be identified. Even if it proves impractical to comfirm most LSST-identified exoplanet candidates via traditional followup observations, it may be possible to derive bulk statistical properties for this exoplanet population.

This paper is the first in a series that will model the predicted properties of the LSST data set to examine:
\begin{itemize}
  \setlength{\itemsep}{0pt}
  \setlength{\parskip}{0pt}
  \setlength{\parsep}{0pt}
  \item What tools can be used to extract exoplanet transit candidates from the LSST light curves?
  \item What will be the dominant types of false positives?
  \item How can the likelihood of detecting these false positives be mitigated, using either the LSST data themselves or follow-up observations?
  \item What new populations of stars can be probed for planets with LSST that have not yet been searched thoroughly?
  \item What do we predict will be the number and parameter distribution of exoplanets detectable with the LSST?
\end{itemize}
This sort of inquiry has been partially addressed by other papers. The LSST Science Book (\citet{LSSTScience2009}, see Section 8.11), and separately \citet{Beatty2008}, examine how many transiting planets might be detected by LSST. We intend to build on these initial examinations by utilizing detailed LSST cadence and photometric performance attributes applied to different types of stars, as well as incorporating updated information about the intrinsic distribution of planet sizes and periods. In later papers will will also include realistic models of stellar binarity and variability, as well as other false positive sources, and investigate improvements to existing transit-detection algorithms customized for the LSST cadence and multiband data.

In this first paper, we present our statistical framework, and we demonstrate that certain types of interesting transiting planets should be detectable in LSST data. We describe our method for modeling LSST light curves in \S \ref{LSSTparam} and describe simple models for simulating and recovering transiting exoplanets in the light curves (\S \ref{SimulatedLC}). In \S \ref{lightcurves} we present the results of this study, in which we demonstrate that a number of scientifically interesting planet/star configurations should be readily detectable in LSST data, even with the use of standard transit-search routines. We close by discussing the key challenges of this project in \S \ref{concluding}.

\section{Assumed LSST Parameters}\label{LSSTparam}
The Large Synoptic Survey Telescope (LSST) is an 8.4-meter telescope being constructed in Cerro Pachon, Chile, with first light planned in 2020 \citep{Ivezic2008}. LSST will survey approximately 30,000 square degrees of the sky repeatedly over the course of 10 years. Observation of a field will consist of two consecutive exposures of 15 seconds each, hereafter refered to as 'visits'. For the majority of the sky, each field will have $\sim$ 1000 visits. This will result in approximately 1 billion light curves that are sparsely sampled over time. LSST will also have a few selected fields that will be observed at a much higher frequency, referred to as deep-drilling fields. Each of these fields will represent 1\% of the total number of LSST observations and will receive $\sim$ 10,000 visits over LSST's scheduled operation. For both cadences (regular cadence and deep-drilling), these observations will be distributed among 6 different wavelength bands, \emph{ugrizy}. LSST's sensitivity will cover a magnitude range from $m\sim$ 16 down to $m\sim$ 25. In this section we describe the cadence and noise parameters that we adopt in our simulated light curves.

\subsection{Cadence Model}
The LSST team has developed the LSST Operations Simulator (OPSIM) software tool that includes models of the hardware and software performance, site conditions, and cadence as a function of sky position \citep{Delgado2014}. For our purposes, OPSIM produces a list of observation times with field ID, band, and limiting magnitude. For this paper we use the OPSIM v2.3.2, run 3.61 results\footnote{Available at \url{https://www.lsstcorp.org/?q=opsim/home}} for two fields --- one representative of a standard field, and one representative of a deep-drilling field --- in order to examine the effect of the two LSST cadences on the observed lightcurves.

\begin{figure}[!htb]
  \begin{center}
    \begin{subfigure}[b]{0.45\textwidth}
      \includegraphics[width=\textwidth]{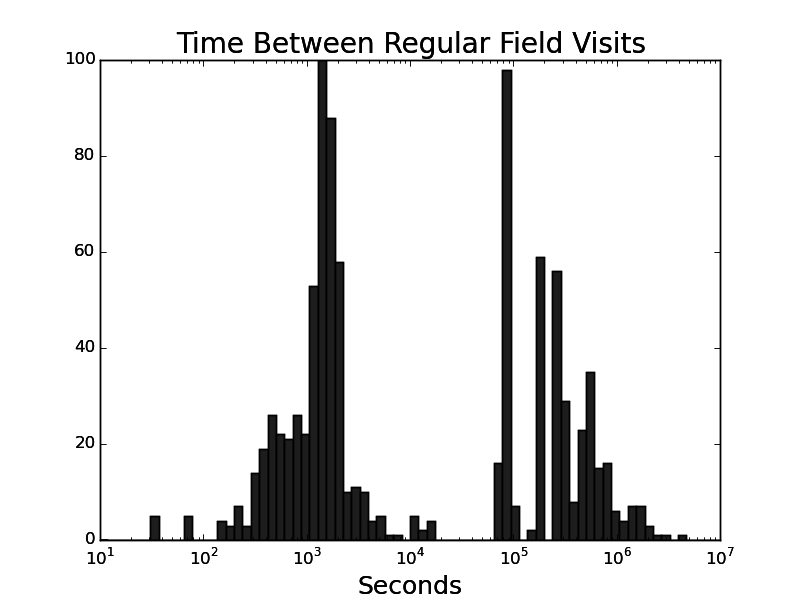}
    \end{subfigure}
    \begin{subfigure}[b]{0.45\textwidth}
      \includegraphics[width=\textwidth]{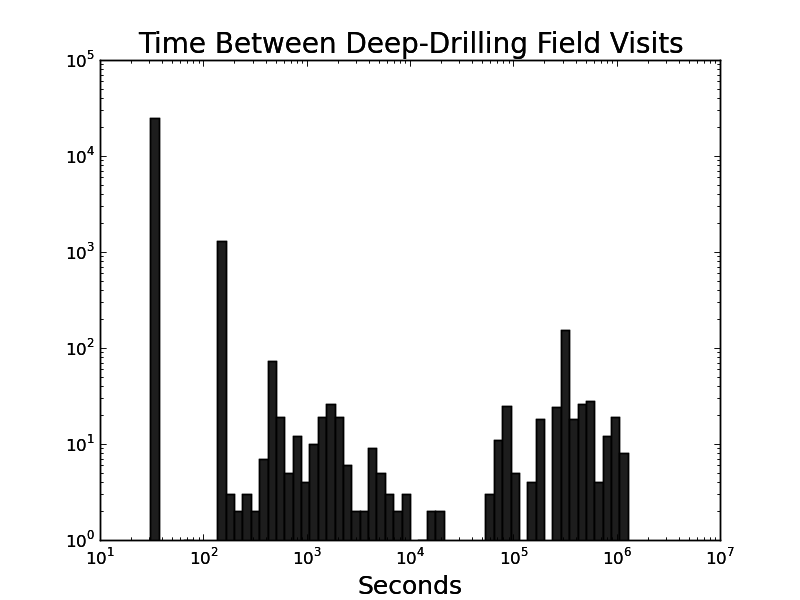}
    \end{subfigure}
  \end{center}
  \caption{Time between beginning of subsequent exposures for a given field, for both regular cadence (left) and deep-drilling cadence (right). Note that a large fraction of the deep-drilling observations are taken in immediate succession, resulting in the largest peak at $\sim$ 34 seconds, corresponding to the combined LSST exposure and readout times.}
  \label{fielddelta}
\end{figure}

In Figure~\ref{fielddelta}, we show the time between consecutive exposures for both fields. In both cases, we have excluded nine intervals that represent the gaps between seasons of observing, which are on the order of 200 days. For the deep drilling field, it is worth noting that because the observing schedules focus on hour-long blocks of constant observation, there is a large peak at $\sim$ 35 seconds representing exposure time plus readout time for consecutive observations of the same field. This then represents the minimum interval between most observations.

\subsection{Noise Model}
\begin{figure}[!htb]
  \begin{center}
    \includegraphics[scale=0.5]{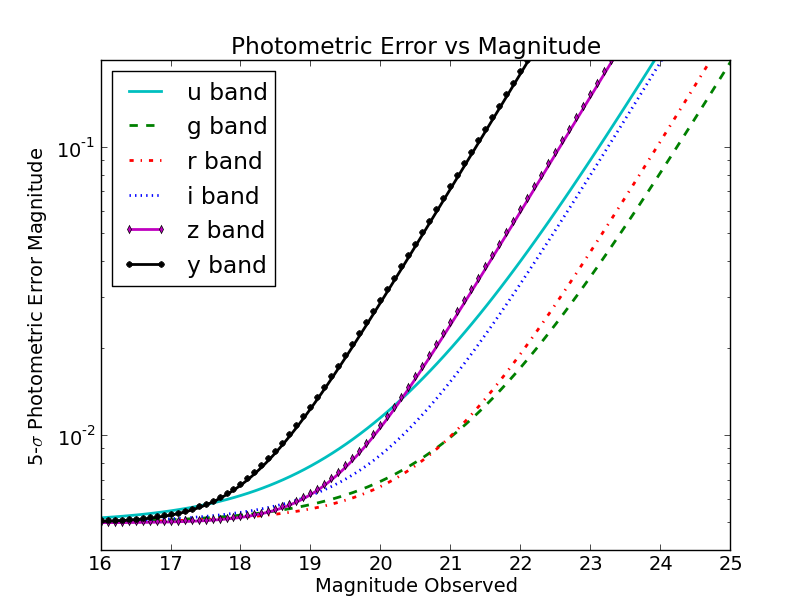}
  \end{center}
  \caption{Noise model at zenith for LSST as a function of stellar magnitude in all observed bands. From \protect\citet{Ivezic2008}}
  \label{noisemodel}
\end{figure}

The photometric noise in simulated LSST light curve comprises two components, a systematic noise floor and random photon noise. LSST's systems are designed to maintain low systematic error with $\sigma_{sys} <$ 0.005 mag. For purposes of this paper, the systematic error is treated conservatively as $\sigma_{sys}$ = 0.005 mag. The random photon noise, dependent on the band being used, varies with changes in atmospheric extinction, stellar magnitude, and seeing conditions. \citet{Ivezic2008} define the random photometric error with the following equation:
\begin{equation} \label{photnoise}
\sigma^{2}_{rand} = (0.04 - \gamma)x + {\gamma}x^{2} \makebox[0pt][l]{\hspace{3cm} $[\mathrm{mag}^{2}]$}
\end{equation}
where \emph{$\gamma$} is a band-specific parameter, and $x = 10^{(m-m_{5})}$. Here, \emph{m} is the band-specific apparent magnitude and \emph{$m_{5}$} is the $5\sigma$ limiting magnitude for point sources. The quantity \emph{$m_{5}$} includes both band-specific parameters and factors such as seeing conditions, air mass, and exposure time. We can then express the total noise for a single visit:
\begin{equation} \label{totnoise}
\sigma^{2} = \sigma^{2}_{sys} + \sigma^{2}_{rand}
\end{equation}
Figure~\ref{noisemodel} shows the total noise for a single LSST visit to any field at the zenith, adopting \emph{$m_{5}$} and \emph{$\gamma$} from \citet{Ivezic2008}. As transits of large exoplanets frequently have depths on the order of 0.01 mag, Figure~\ref{noisemodel} shows that for apparent magnitude $\sim 16$ down to as far as $\sim 20$, for some bands the photometric noise is less than the depth that would be expected of a transit event. This represents a much fainter apparent magnitude range with this level of photometric precision than most previous transiting planet searches have explored.

\section{Simulated LSST Light Curves of Transits}\label{SimulatedLC}
\subsection{Light Curve Construction}

We create synthetic light curves to analyze the ability to detect transit signatures. In \S \ref{lightcurves} we define several star/planet configurations for simulated light curves, specifying for each the stellar mass, planetary radius, orbital period, and distance from Earth. Stellar masses are converted to spectral types by interpolating the relations from \citet{Cox2000}. Based on spectral type, we obtain absolute magnitudes in the \emph{ugriz} bands from \citet{Covey2007}. While there is still much uncertainty about the specific parameters of the \emph{y}-band filter that LSST will eventually use, for this work we use a \emph{y}-band defined by \citet{Hodgkin2009}.

Light curve generation itself requires orbital period, as well as transit depth and duration as determined by the geometry of the system, and the apparent magnitude of the host star in all bands. \emph{Kepler} has shown that while stars do have a wide range of intrinsic stellar variability, most will have variability of less than 1\% \citep{Basri2011}. This intrinsic noise will be smaller than both the noise present in the observations and the transit depths we focus on, so we do not include that effect in these light curves. Transits are modeled as boxcars and when applied to the apparent magnitude of the star, allow us to create a noiseless light curve of the star that features the transit. We then add the photometric noise to the light curve, modeled as Gaussian noise, and dependent on apparent magnitude and band (See Figure~\ref{noisemodel}) to create light curves similar to what LSST will observe.

\begin{figure}[!htb]
  \begin{center}
    \includegraphics[scale=0.6]{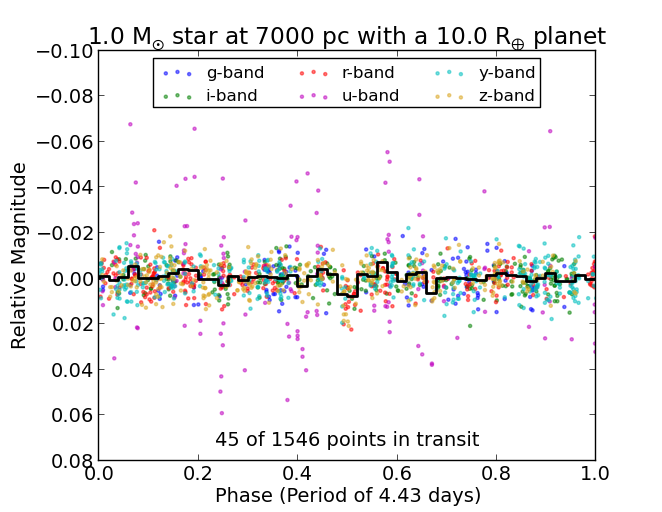}
  \end{center}
  \caption{A sample six-band light curve, demonstrating the ability to combine all bands into a single phased light curve. The black line shows the binned data for the light curve.}
  \label{6band}
\end{figure}

The resulting light curve can be represented in all six bands. Limb darkening is a second-order effect that will tend to decrease the transit depth at blue wavelengths relative to a simple box-shaped transit. However, to first order the transit depth will be independent of wavelength, so in this initial exploration we treat the transits as achromatic boxcar events. With this simplification we can median-subtract each light curve and combine them to create one unified light curve for enhanced transit detection. An example of a six-band light curve is shown in Figure~\ref{6band}. It is already noticeable in this case that the \emph{u}-band will have the largest noise for the stellar populations in which we are most interested.

The multiband nature of the observations will also provide additional information to characterize the light curves. Since exoplanet transits would be expected to be the same depth in all wavelengths, signals from eclipsing binary stars can be eliminated as planet candidates if we observe transit depths that are dependent on wavelength. Our intention in this paper is not to provide a rigorously developed number for expected exoplanet detections, or to address distinguishing planetary transits from other signals, but rather to demonstrate that exoplanets will be theoretically detectable, and that this is a worthwhile consideration as the LSST project moves forward.

\subsection{Period Recovery}
We examine the recoverability of transits by searching for periodicity using the Box-fitting Least Squares (BLS) algorithm \citep{Kovacs2002}\footnote{We use the implementation of the BLS algorithm that is part of the VARTOOLS package\protect\citep{Hartman2008}}. The BLS algorithm has been shown to be extremely powerful for detecting transits in high cadence data sets \citep{Enoch2012}. While LSST will not be a high cadence survey, we use this algorithm as an initial test of LSST’s capability for recovering transit signals. In order to examine the significance of these detections, we also measure the liklihood of having a peak of equal strength in the BLS periodogram by random chance, or due to some feature of the cadence, to measure a false alarm probability (FAP). To carry out this check, for each light curve we randomly rearrange the magnitude and time information, and then reapply BLS. We conduct that operation 1000 times for each light curve, and record the highest BLS peak in each case.  The resulting data sets allows us to specify the 1\% and 0.1\% false alarm probabilities for BLS.

\section{Results: Examples of Simulated Light Curves for Different Types of Transits} \label{lightcurves}

We begin examining detectability by creating simulated light curves (\S \ref{SimulatedLC}) to examine what exoplanet transits will look like for several different cases with stellar parameters outlined in Table~\ref{table:StellarStat}. These are intended to exemplify well-studied populations of exoplanets as well as some populations that could prove very interesting but have not been a focus of planet searches thus far.

\begin{table}[!htb]
\caption{Photometric Properties of Example Systems}
\label{table:StellarStat}
  \begin{center}
  \small
  \begin{tabular}{rrrrrrrr}
    \hline
    & & \multicolumn{6}{c}{Band} \\ \hline
    Scenario &  & \textit{$m_u$} & \textit{$m_g$} & \textit{$m_r$} & \textit{$m_i$} & \textit{$m_z$} & \textit{$m_y$} \\ \hline\hline
    \multirow{2}{*}{G-dwarf at 7000 pc} & Apparent Magnitude & 21.433 & 20.066 & 18.699 & 18.540 & 18.500 & 17.913 \\ 
    & Standard Deviation & 0.023 & 0.007 & 0.006 & 0.006 & 0.006 & 0.007 \\
    \hline
    \multirow{2}{*}{K-dwarf at 2000 pc} & Apparent Magnitude & 23.619 & 21.091 & 18.526 & 17.980 & 17.675 & 17.265 \\ 
    & Standard Deviation & 0.149 & 0.011 & 0.005 & 0.005 & 0.005 & 0.006 \\
    \hline
    \multirow{2}{*}{M-dwarf at 400 pc} & Apparent Magnitude & 24.424 & 21.466 & 18.530 & 17.150 & 16.428 & 15.704 \\ 
    & Standard Deviation & 0.286 & 0.013 & 0.005 & 0.005 & 0.005 & 0.005 \\
    \hline
    \multirow{2}{*}{G-dwarf at 50 kpc} & Apparent Magnitude & 25.658 & 24.334 & 22.968 & 22.807 & 22.770 & 22.183 \\ 
    & Standard Deviation & 1.029 & 0.109 & 0.043 & 0.068 & 0.122 & 0.214 \\
    \hline
  \end{tabular}
  \end{center}
  {Absolute magnitudes for \emph{ugriz} bands from \citet{Covey2007} and \emph{y} band from \citet{Hodgkin2009}, with corresponding photometric noise.}
\end{table}
\subsection{A Hot Jupiter Orbiting a G-dwarf in the Milky Way}
\begin{figure}[!htb]
  \begin{center}
    \begin{subfigure}[b]{0.45\textwidth}
      \includegraphics[width=\textwidth]{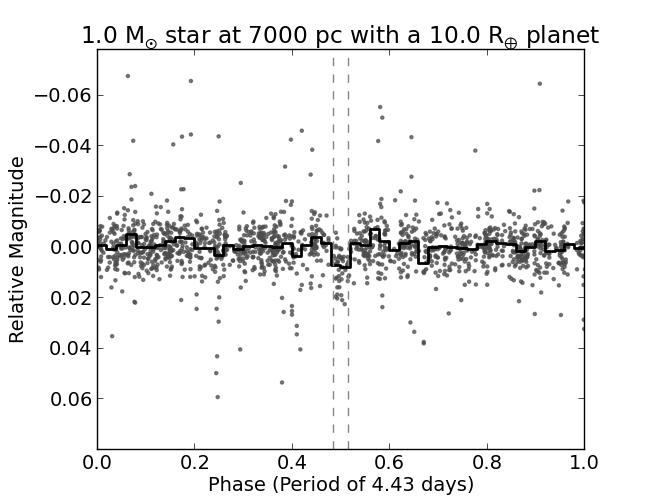}
    \end{subfigure}
    \begin{subfigure}[b]{0.45\textwidth}
      \includegraphics[width=\textwidth]{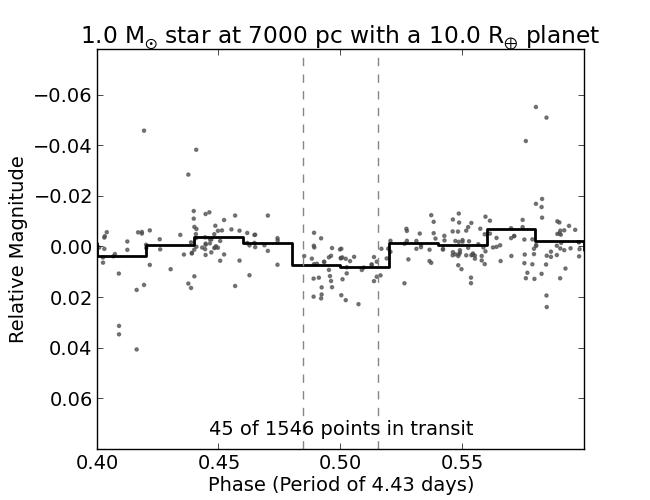}
    \end{subfigure}
    \begin{subfigure}[b]{0.45\textwidth}
      \includegraphics[width=\textwidth]{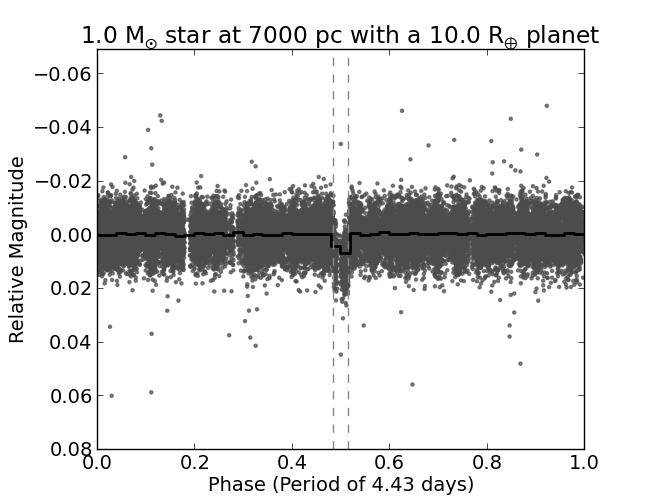}
    \end{subfigure}
    \begin{subfigure}[b]{0.45\textwidth}
      \includegraphics[width=\textwidth]{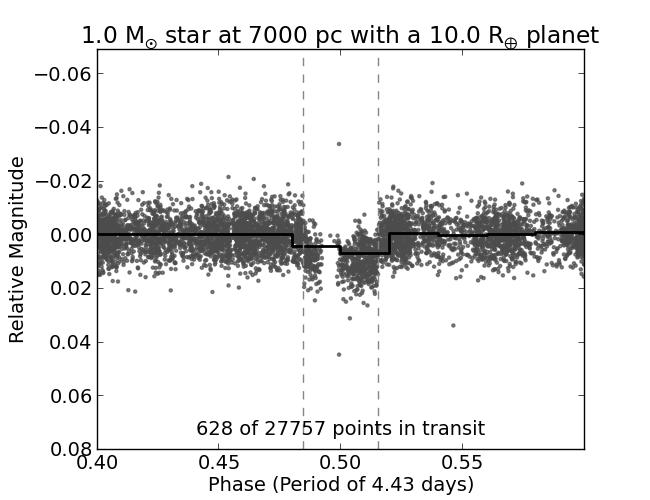}
    \end{subfigure}
  \caption{Light curve for a 10.0 $R_{\oplus}$ planet in a 4.43 day period around a 1.0 $M_{\astrosun}$ star at 7000 pc. The top two plots show a regular LSST field and the bottom two plots show an LSST deep-drilling field. The plots on the left show the full phase, and the plots on the right zoom in on the transit. Black lines are binned data of the light curve.}
  \label{HJlc}
  \end{center}
\end{figure}

The category of transiting exoplanets that have been studied most extensively are the Hot Jupiters, Jovian-mass planets orbiting sun-like stars with periods on the order of days. Here, we simulate a light curve for a 10.0 $R_{\oplus}$ planet orbiting at 4.43 days around a 1.0 $M_{\astrosun}$ star. We have set the star at a distance of 7000 pc, chosen so that the apparent magnitude of the host star is within the sensitivity range of LSST in all bands, while also representing a distance inaccessible to most current exoplanet surveys. The resultant light curve is then phase-folded on the input period in order to generate the phased light curve displayed in Figure~\ref{HJlc}. In this case, there is a very noticeable drop in the star's brightness of 0.009 mag, showing that LSST can at least detect such a transit strong enough in the six-band lightcurve to be visually identifiable.

\begin{figure}[!htb]
  \begin{center}
    \begin{subfigure}[b]{0.45\textwidth}
      \includegraphics[width=\textwidth]{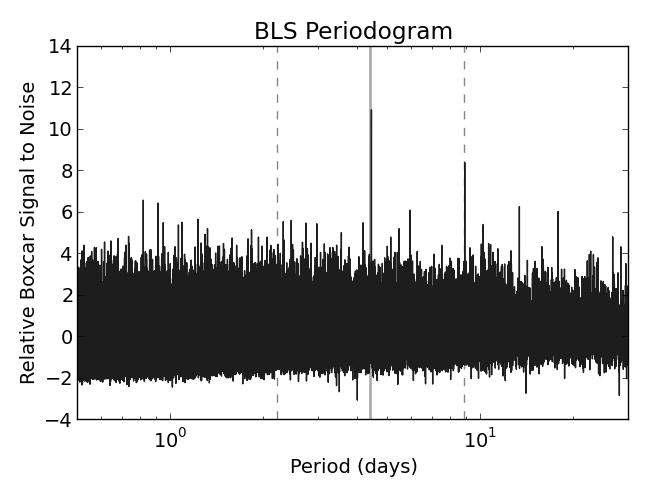}
    \end{subfigure}
    \begin{subfigure}[b]{0.45\textwidth}
      \includegraphics[width=\textwidth]{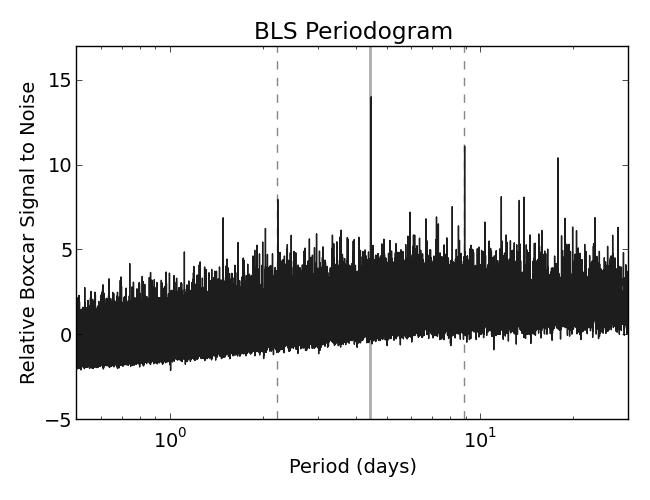}
    \end{subfigure}
  \caption{BLS periodogram for a Hot Jupiter in a standard field on the left, and a deep drilling field on the right. The lighter grey line marks the actual period of 4.43 days, while the other two dashed lines mark twice and half the true period.}
  \label{HJbls}
  \end{center}
\end{figure}

For the Hot Jupiter case, we provide the BLS periodograms for both a regular LSST field and a deep-drilling field in Figure~\ref{HJbls}. We also mark the true period on the periodogram, as well as the aliases at half and double the period. In both cases, the 4.43 day period is the highest peak in the periodogram, providing a qualitative reassurance that that the periods visible in Figure~\ref{HJlc} could be recovered had the value not been known. We find that in 1000 light curves there are no peaks of equal or greater height, from which we determine the false positive rates for Hot Jupiters at both cadences to be $<$0.1\%.

\subsection{A Hot Neptune Orbiting a K-dwarf in the Milky Way}

Next we examine a Hot Neptune. Representing a regime between very large Jovian planets and smaller terrestrial planets (radii 3-4 $R_{\oplus}$) that has been richly populated by \emph{Kepler}. Here we place a 4.0 $R_{\oplus}$ planet in a 7.3 day period around a 0.6 $M_{\astrosun}$ star at 2000 pc.

\begin{figure}[!htb]
  \begin{center}
    \begin{subfigure}[b]{0.45\textwidth}
      \includegraphics[width=\textwidth]{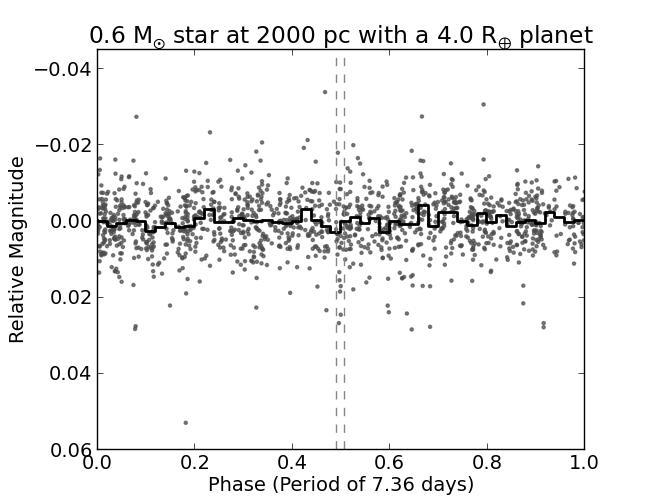}
    \end{subfigure}
    \begin{subfigure}[b]{0.45\textwidth}
      \includegraphics[width=\textwidth]{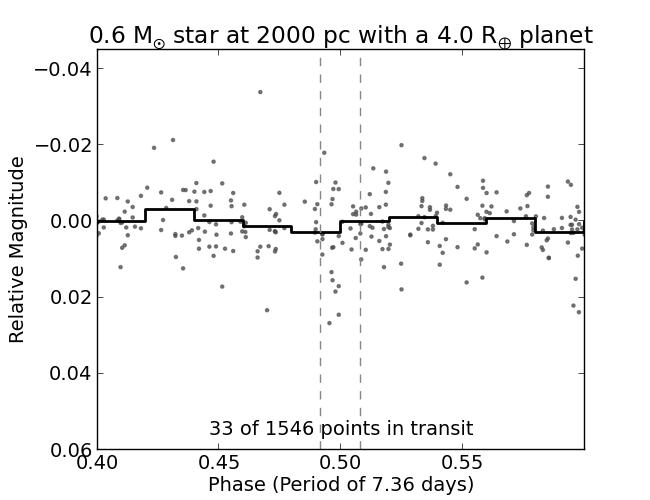}
    \end{subfigure}
    \begin{subfigure}[b]{0.45\textwidth}
      \includegraphics[width=\textwidth]{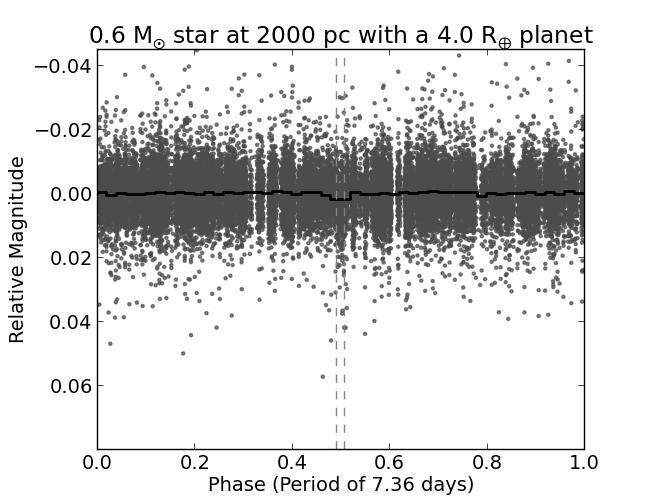}
    \end{subfigure}
    \begin{subfigure}[b]{0.45\textwidth}
      \includegraphics[width=\textwidth]{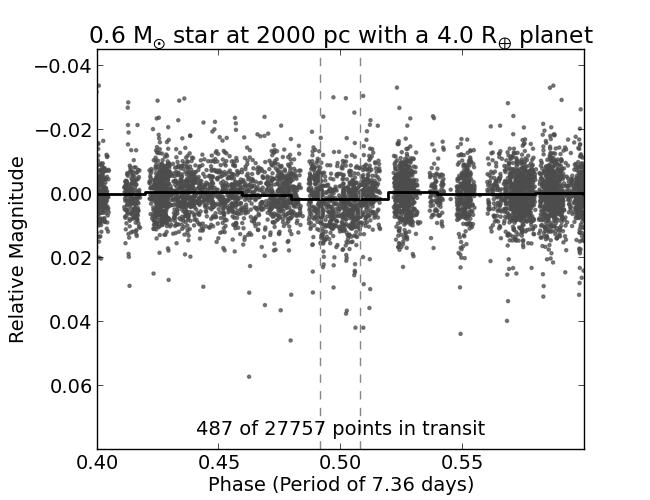}
    \end{subfigure}
  \caption{A 4.0 $R_{\oplus}$ planet in a 7.36 day period around a 0.6 $M_{\astrosun}$ star at 2000 pc.  The top two plots show a regular LSST field and the bottom two plots show an LSST deep-drilling field. The plots on the left show the full phase of the planet, and the plots on the right show the transit in particular. Black lines are binned data of the light curve.}
  \label{HNlc}
  \end{center}
\end{figure}

Ground-based observations are generally limited in sensitivity to transit depths on the order of 1\%, so we have used a slightly smaller host star such that a Neptune-sized planet will still represent an approximately similar drop in brightness from the Hot Jupiter case. The closer distance is to maintain the constraint that the star's apparent magnitude in all bands will be within the sensitivity range of LSST. As the transit events that we are examining will generally range between 3 mmag and 10 mmag in depth, we chose only to plot bands where the standard deviation is less than 30 mmag. In this light curve the noise in the \emph{u}-band dominates the noise from the other bands, and we only plot the \emph{grizy} points. These light curves are shown in Figure~\ref{HNlc}.

\begin{figure}[!htb]
  \begin{center}
    \begin{subfigure}[b]{0.45\textwidth}
      \includegraphics[width=\textwidth]{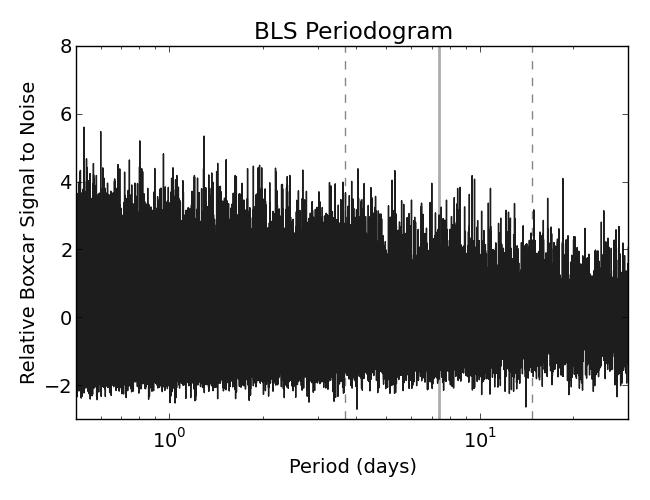}
    \end{subfigure}
    \begin{subfigure}[b]{0.45\textwidth}
      \includegraphics[width=\textwidth]{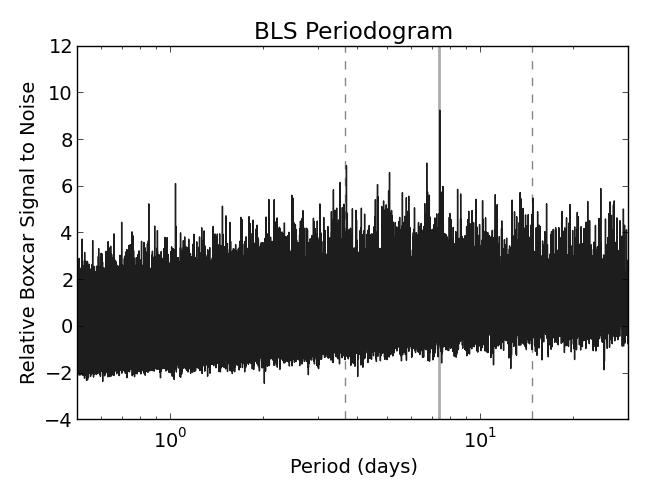}
    \end{subfigure}
  \caption{BLS periodogram for a Hot Neptune in a standard field on the left, and a deep drilling field on the right. The lighter grey line marks the actual period of 7.36 days, while the two dashed lines mark twice and half the true period.}
  \label{HNbls}
  \end{center}
\end{figure}

The transit in Figure~\ref{HNlc} is not as visible as in the case of a Hot Jupiter, but can be discerned in the deep-drilling field where there are enough observations for the transit to be seen despite the noise. We provide the BLS periodograms for both a regular LSST field and a deep-drilling field in Figure~\ref{HNbls}, and indicate the true period. For the regular LSST field BLS does not recover the period and there are no strong discernable features in the periodogram that would correspond to the planet period. However, BLS does recover the input period for the deep-drilling field. This peak is higher than all but one BLS period from the 1000 permutated light curves, and a false positive rate of 0.1\%.

\subsection{A Super-Earth Around an M-dwarf in the Milky Way}
\begin{figure}[!htb]
  \begin{center}
    \begin{subfigure}[b]{0.45\textwidth}
      \includegraphics[width=\textwidth]{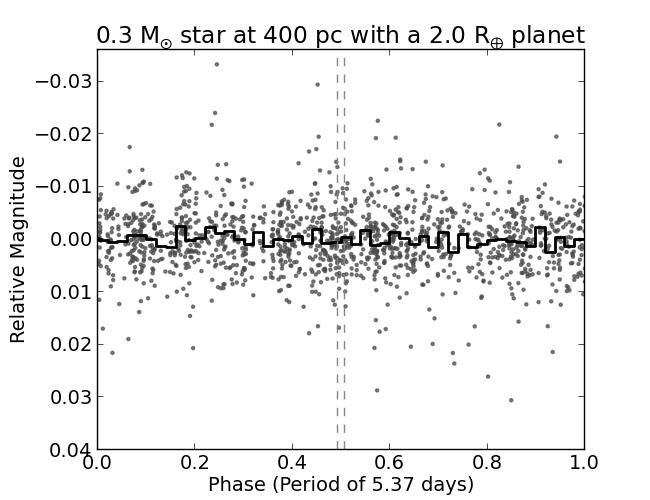}
    \end{subfigure}
    \begin{subfigure}[b]{0.45\textwidth}
      \includegraphics[width=\textwidth]{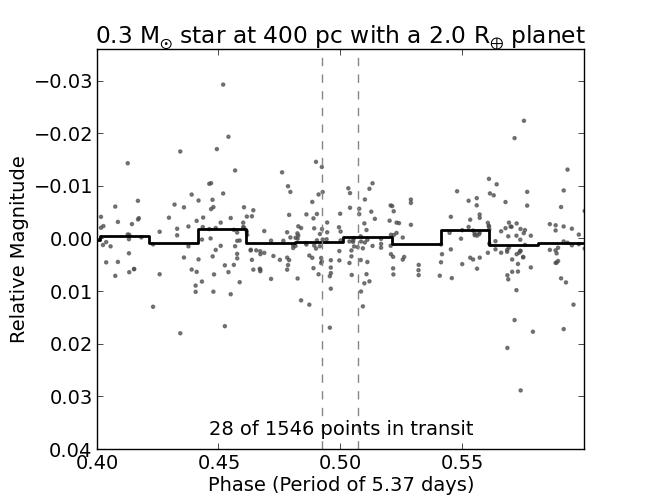}
    \end{subfigure}
    \begin{subfigure}[b]{0.45\textwidth}
      \includegraphics[width=\textwidth]{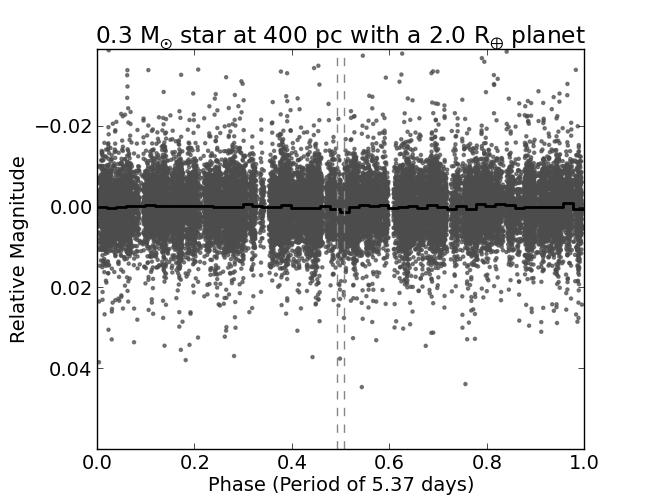}
    \end{subfigure}
    \begin{subfigure}[b]{0.45\textwidth}
      \includegraphics[width=\textwidth]{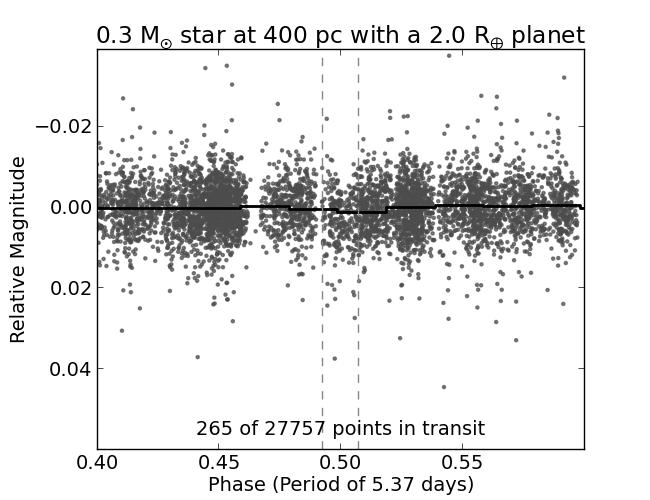}
    \end{subfigure}
  \caption{A 2.0 $R_{\oplus}$ planet in a 5.37 day period around a 0.3 $M_{\astrosun}$ star at 400 pc. The top two plots show a regular LSST field and the bottom two plots show an LSST deep-drilling field. The plots on the left show the full phase of the planet, and the plots on the right show the transit in particular. Black lines are binned data of the light curve.}
  \label{SElc}
  \end{center}
\end{figure}

\begin{figure}[!htb]
  \begin{center}
    \begin{subfigure}[b]{0.45\textwidth}
      \includegraphics[width=\textwidth]{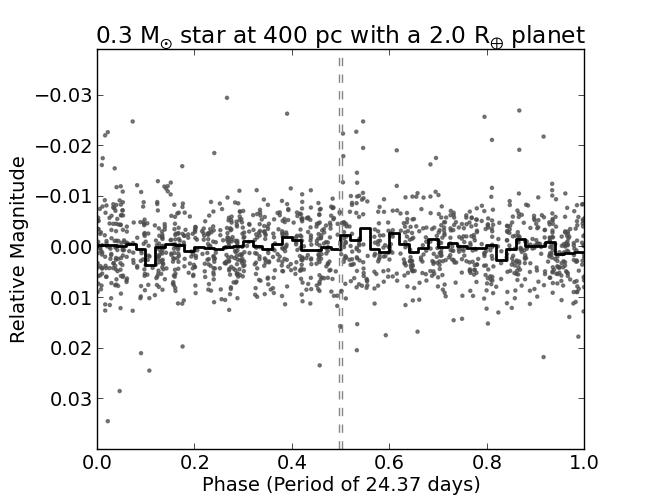}
    \end{subfigure}
    \begin{subfigure}[b]{0.45\textwidth}
      \includegraphics[width=\textwidth]{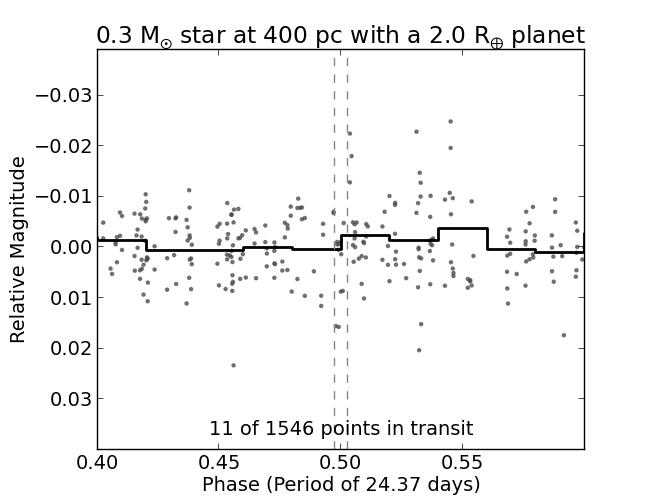}
    \end{subfigure}
    \begin{subfigure}[b]{0.45\textwidth}
      \includegraphics[width=\textwidth]{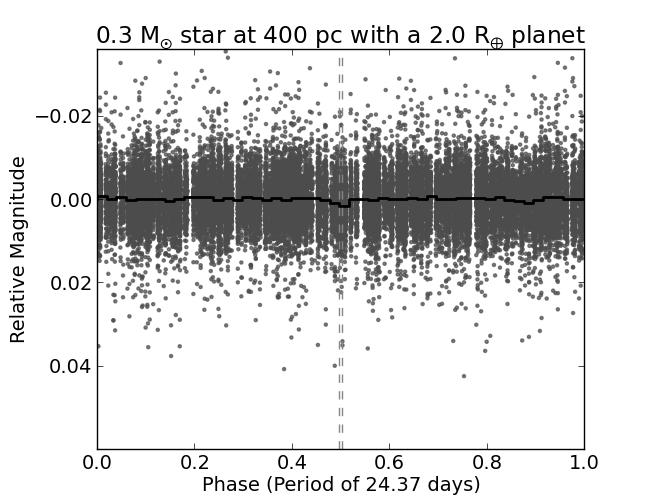}
    \end{subfigure}
    \begin{subfigure}[b]{0.45\textwidth}
      \includegraphics[width=\textwidth]{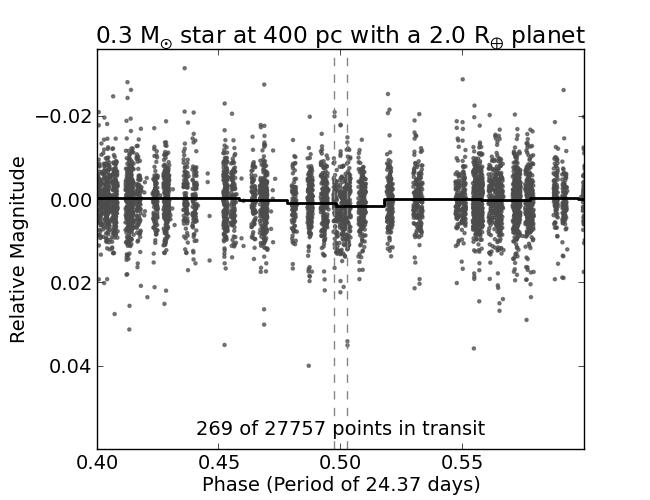}
    \end{subfigure}
  \caption{A 2.0 $R_{\oplus}$ planet in a 24.37 day period around a 0.3 $M_{\astrosun}$ star at 400 pc. The top two plots show a regular LSST field and the bottom two plots show an LSST deep-drilling field. The plots on the left show the full phase of the planet, and the plots on the right show the transit in particular. Black lines are binned data of the light curve.}
  \label{SEhlc}
  \end{center}
\end{figure}

For our final Milky Way case, we pick a 2.0 $R_{\oplus}$ planet in a 5.37 day period around a 0.3 $M_{\astrosun}$ star at 400 pc. The distance has again been adjusted so that the apparent magnitude of the red dwarf is approximately still within LSST's sensitivity range. As in the previous case, the \emph{u}-band is omitted. In Figure~\ref{SElc} we show the light curves for the transit in both a standard field and a regular field. While in the standard field the transit is not visually detectable, in the deep-drilling field a small but slightly noticeable transit event is visible in the 5 remaining bands. With what appears to be a detectable signal in the deep-drilling field, we further explore a red dwarf host star with a Super-Earth in a larger period, moving the planet out to a period of 24.37 days. Our motivation in choosing this period is that for a red dwarf of this mass, a planet in a $\sim$25 day period will be in the habitable zone, and this planet's system environment would be similar to that calculated for some planets in the Gliese 667C system \citep{Gregory2012}. The light curve for this case is presented in Figure~\ref{SEhlc}.


\begin{figure}[!htb]
  \begin{center}
    \begin{subfigure}[b]{0.45\textwidth}
      \includegraphics[width=\textwidth]{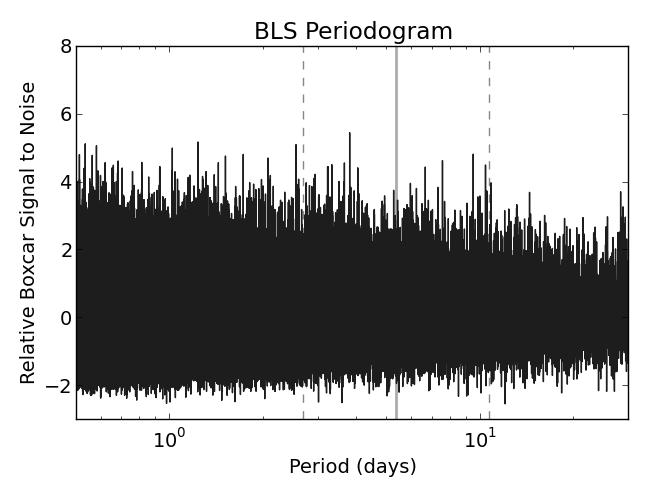}
    \end{subfigure}
    \begin{subfigure}[b]{0.45\textwidth}
      \includegraphics[width=\textwidth]{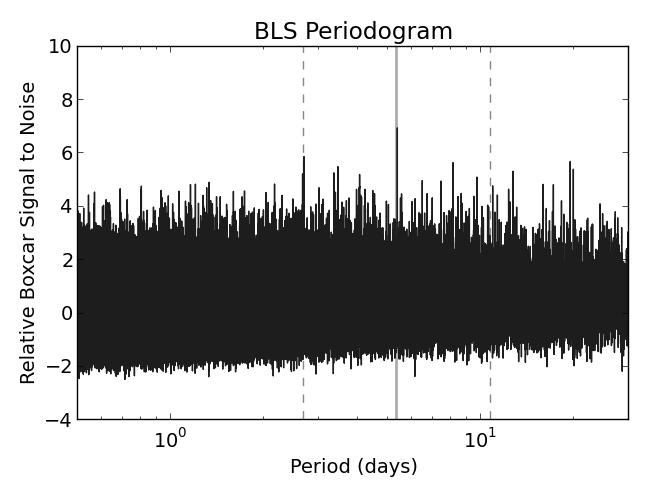}
    \end{subfigure}
    \begin{subfigure}[b]{0.45\textwidth}
      \includegraphics[width=\textwidth]{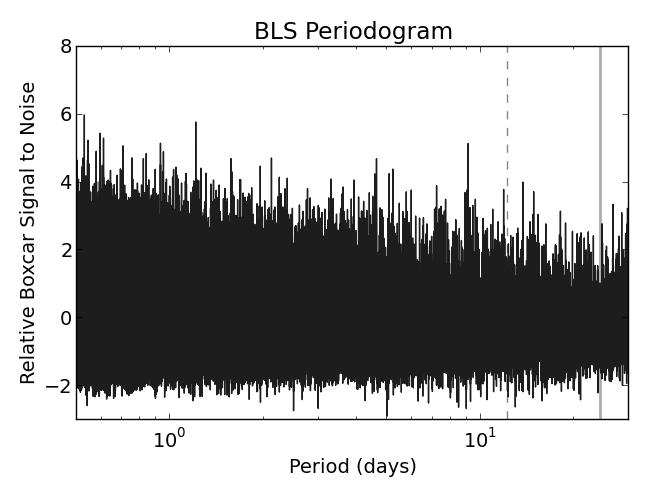}
    \end{subfigure}
    \begin{subfigure}[b]{0.45\textwidth}
      \includegraphics[width=\textwidth]{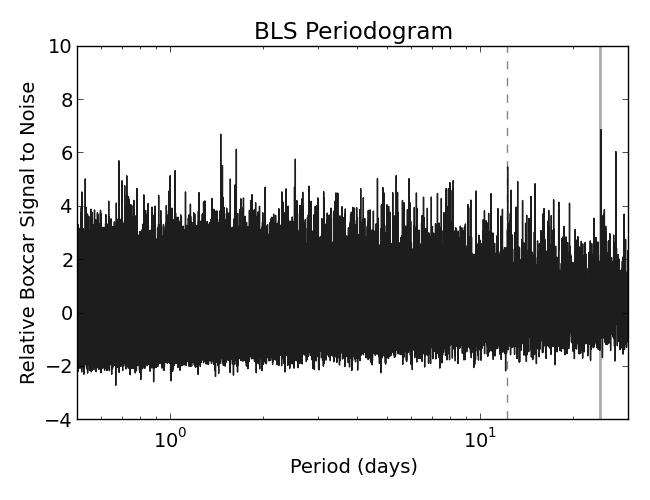}
    \end{subfigure}
  \caption{BLS periodograms for a Super-Earth in a regular field on the left and in a deep-drilling field on the right. The 5.37-day period is on the top and the 24.37-day period is the bottom. The lighter grey line marks the actual period for each scenario.}
  \label{SEbls}
  \end{center}
\end{figure}

Figure~\ref{SEbls} shows the BLS periodograms for both the 5.37 and 24.37 day periods, and at both cadences. In both cases the input period corresponds to the highest peak in the periodogram\footnote{We note that in this case, the BLS SN statistic identified the top peak as being one of the secondary peaks in the periodogram (Fig.~\ref{SEbls}) even though it is clear that the input period does coincide with the highest peak in the periodogram. We will examine the relevant differences between the various peak identification statistics and the BLS periodogram in a future paper.}. In the 1000 permutations of the simulated habitable Super-Earth light curve in a deep-drilling field, we find that there are 390 higher peaks, corresponding to a false positive rate of 19.5\%.  Looking at these periodograms, however, that there is still potentially useful information present in the BLS results. While there appears to be no compelling features in the two lefthand figures that represent the two periods in regular fields, there are notable features in the deep-drilling fields on the right. For the 5.37 day period, the two highest peaks are the initial period and half the period. Similarly, for the 24.37 day period the initial period is one of the highest peaks, and there is a notable peak at half the period as well. While these period recoveries lack the rigor of being the top peak recovered by BLS, this does provide an indication that the transits could still be more reliably recovered in deep-drilling fields with the application of additional methods.

\subsection{A Hot Jupiter in the Large Magellanic Cloud}

\begin{figure}[!htb]
  \begin{center}
    \begin{subfigure}[b]{0.45\textwidth}
      \includegraphics[width=\textwidth]{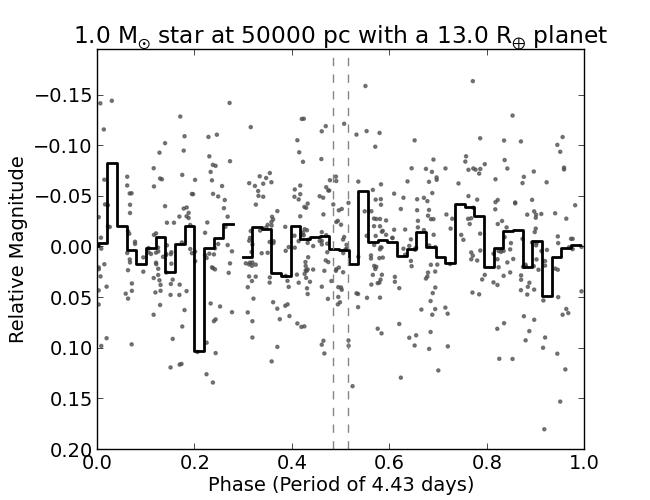}
    \end{subfigure}
    \begin{subfigure}[b]{0.45\textwidth}
      \includegraphics[width=\textwidth]{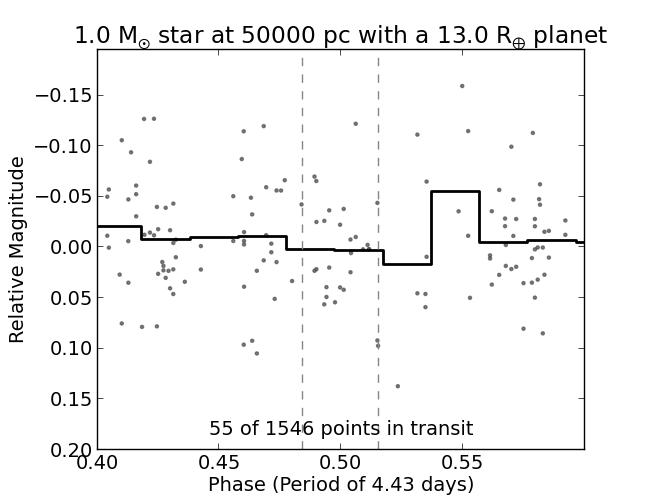}
    \end{subfigure}
    \begin{subfigure}[b]{0.45\textwidth}
      \includegraphics[width=\textwidth]{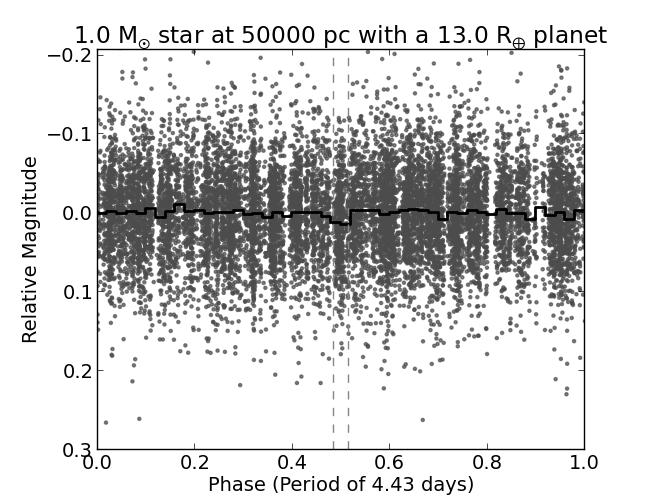}
    \end{subfigure}
    \begin{subfigure}[b]{0.45\textwidth}
      \includegraphics[width=\textwidth]{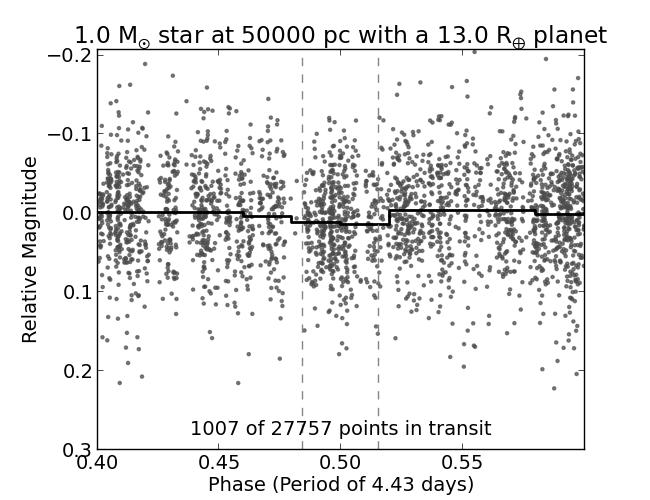}
    \end{subfigure}
  \caption{A 13.0 $R_{\oplus}$ planet in a 4.43 day period around a 1.0 $M_{\astrosun}$ star at 50,000 pc. The top two plots show a regular LSST field and the bottom two plots show an LSST deep-drilling field. The plots on the left show the full phase of the planet, and the plots on the right show the transit in particular. Black lines are binned data of the light curve.}
  \label{LMC1lc}
  \end{center}
\end{figure}

\begin{figure}[!htb]
  \begin{center}
    \begin{subfigure}[b]{0.45\textwidth}
      \includegraphics[width=\textwidth]{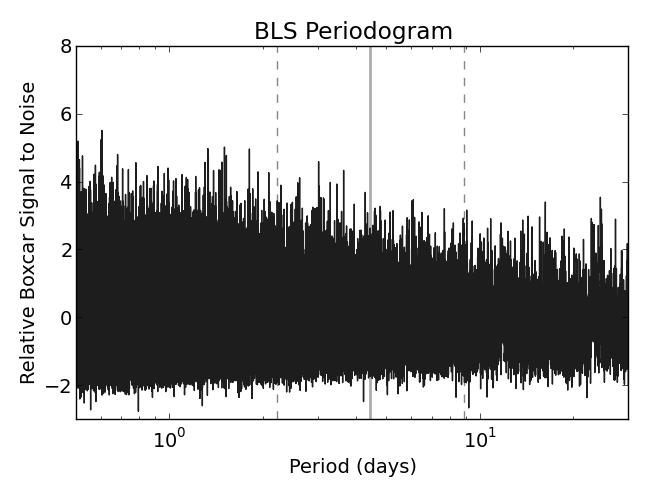}
    \end{subfigure}
    \begin{subfigure}[b]{0.45\textwidth}
      \includegraphics[width=\textwidth]{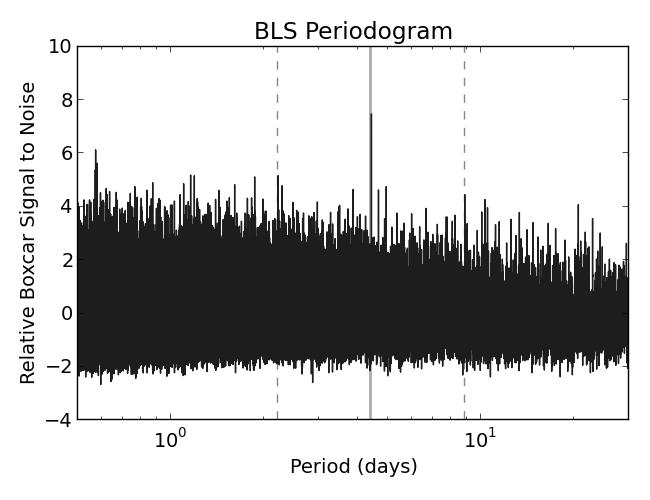}
    \end{subfigure}
  \caption{BLS periodogram for a Hot Jupiter orbiting a star in the LMC in a regular field on the left and a deep-drilling field on the right. The lighter grey line marks the actual period.}
  \label{LMCHJbls}
  \end{center}
\end{figure}

LSST will also provide a unique opportunity to look for exoplanets outside of the Milky Way. Fields that have been proposed for LSST to observe at higher cadence include the Large Magellanic Cloud \citep{Szkody2011}. Based on the faintness that LSST will be capable of observing, we demonstrate here that LSST should have the potential to be sensitive to Hot Jupiters transiting stars located in the Large Magellanic Cloud. For this example, we look at a slightly larger Hot Jupiter than before, here a 13 $R_{\oplus}$ planet with a 4.4 day period orbiting a 1.0 $M_{\astrosun}$ star at 50,000 parsecs from Earth in Figure~\ref{LMC1lc}. As in the previous two examples, the \emph{u}-band data has been omitted, but at this faintness the noise is sufficiently large that we also ignore the  \emph{gzy}-bands and only plot the \emph{ri}-bands. While we are unable to detect the transit in the regular cadence light curve, we can detect it in the deep drilling light curve, as shown in Figure~\ref{LMCHJbls}.

The entire set of deep-drilling fields has yet to be chosen, although the first four deep-drilling fields to be selected from LSST white papers have been intended for observations of distant galaxies \citep{Gawiser2011, Ferguson2011}. However, other deep-drilling fields have also been suggested with the intention of focusing on the Large and Small Magellanic Clouds \citep{Szkody2011}. These fields for the LMC and SMC would have an observing schedule which limits the filters used, and is proposed to only use \emph{gr} band observations. This would be consistent with the limitations on photometric precision in the final light curves, which do not include several bands that we omitted due to large noise, and may result in light curves with much better photometric precision.

For a Hot Jupiter in the Large Magellanic Cloud, we provide the BLS periodograms for both a regular LSST field and a deep-drilling field in Figure~\ref{LMCHJbls}. BLS does not recover the period in a regular field, nor do there seem to be any notable features in the periodogram, however we do recover the period in the deep-drilling field with a peak height that indicates a false positive rate of less than 0.1\%. This result demonstrates that LSST could conceivably detect a planet candidate outside the Milky Way. While it is worth noting that at the distance of the LMC it would be very difficult to confirm an exoplanetary candidate, this would certainly be a novel discovery and give some broader indication of planet formation processes in other environments.

\section{Discussion}\label{concluding}

We have demonstrated that using existing algorithms, LSST will have the sensitivity to detect exoplanet transits for an array of exoplanets and host stars, and at varying distances. These results are summarized in Table~\ref{table:Detections}. However, these capabilities will be greatly shaped by the selection process for the remaining deep drilling fields, as the selection of the LMC or SMC for deep drilling will greatly increase the ability to get visually identifiable transits in those light curves.

\begin{table}[!htb]
\caption{Scenario Detection}
\label{table:Detections}
  \begin{center}
  \small
  \begin{tabular}{cccccc}
    \hline
    & \multicolumn{5}{c}{Scenario} \\\cline{2-6}
    & G-dwarf at 7000 pc & K-dwarf at 2000 pc & M-dwarf at 400 pc & M-dwarf at 400 pc & G-dwarf at 50 kpc \\
    Cadence & Hot Jupiter & Hot Neptune & Super-Earth (4.4 d) & Super-Earth (25.3 d) & Hot Jupiter \\
    \hline\hline
    Standard & Yes & No & No & No & No \\
    False Positive Probability & $<10^{-3}$ & -- & -- & -- & --\\
    Deep-Drilling & Yes & Yes & No & No & Yes \\
    False Positive Probability & $<10^{-3}$ & $10^{-3}$ & -- & -- & $<10^{-3}$\\
    \hline
  \end{tabular}
  {Table~\ref{table:Detections} lists whether or not the best period from VARTOOLS BLS algorithm is consistent with the input period for each of the scenarios described in \S \ref{lightcurves} (and listed at the top of each column). The rows represent a standard cadence field and a deep-drilling field. Also included is the false positive rate when the period was correctly recovered by BLS.}
  \end{center}
\end{table}

For the case of Hot Jupiters, specifically, we can make a very preliminary approximation of the number of planets that could be recovered by LSST. Restricting our focus to the \emph{g}-band, we find the number of stars between 16th and 22nd magnitude in a one-square-degree field at around 1800 stars, according to TRILEGAL 1.6\footnote{Available at \url{http://stev.oapd.inaf.it/cgi-bin/trilegal}}. There will, additionally, be around 30,000 square degrees observed, for a total of 54 million stars. Of these, we would expect around 0.5\% of dwarf stars to host Hot Jupiters, and the likleihood of a Hot Jupiter to transit its host star is $\sim$ 10\%. This would result in 27000 transiting Hot Jupiters, and recovering just 5\% of these Hot Jupiters would result in over 1000 new Hot Jupiters discovered. While these are very approximate numbers, they do indicate the large potential that LSST possesses.

Through our work thus far, we have treated host stars as having no intrinsic variability, however a more sophisticated model of stellar light curves will be needed going forward. Stellar variation from rotation, pulsation, flaring, and eclipses will need to be included to better synthesize the kinds of light curves that will likely be observed by LSST, and these variations will complicate recovering the signal caused by a transiting planet and be a source of false negatives. Additionally, we will need to incorporate these light curve features into the light curves of stars in the field without planets in order to properly model false positives.

A greater difficulty will be developing methods that can more rigorously detect and recover these transits, and determine the difference between likely exoplanetary candidates and false positives, astrophysical or otherwise. Due to the low cadence of the data, standard methods used for finding periodic signals in light curves are potentially not sufficient to find transits in all circumstances, and newer algorithms will need to be implemented with these constraints in mind. The efficiency of these algorithms will be an important question, both to quantify the usefulness that LSST will have in detecting extrasolar planets, and in determining the best way to structure a search for transiting planets in the LSST data once observations begin in 2020.

There will also be a need to expand this study from the discrete example cases we have set forth here, and apply similar analysis to a more diverse set of systems. That will include expanding our synthetic light curves to represent a distribution of stellar masses and exoplanet radii and periods, as well as factoring in the likelihood that a planet will transit. This will allow our results to look not just at the likelihood for a given planet to be detected, but also to determine how many exoplanets could reasonably be detected among stellar populations such as field stars and star clusters, and will also characterize the variety of planet types that will be detectable using the LSST data.

While we have discussed potential opportunities for exoplanet detection specifically from LSST, these same methods can be applied to other surveys that have a large number of observations over relatively long baselines, but with similarly sparsely-sampled data that will make extracting exoplanetary transits difficult without implementing more novel algorithms. Surveys that would warrant such consideration may include Pan-STARRS \citep{Kaiser2004}, the Catalina Real-Time Sky Survey \citep{Djorgovski2008}, the Palomar Transient Factory \citep{Rau2009}, and GAIA \citep{Jordi2011}. Once our procedures for stellar and exoplanetary population simulation, light curve creation, and application of detection algorithms have been developed for LSST, we will then be able to apply these same techniques to these other surveys to greatly increase the data available to search for exoplanets in varied situations.

\acknowledgments
We acknowledge support through NASA ADAP grant NNX12AE22G and through the Vanderbilt Initiative in Data-intensive Astrophysics.

\bibliographystyle{hapj}
\bibliography{libAAS}

\end{document}